\documentclass[showpacs,amsmath,amssymb,aps,showkeys,floatfix,prd,a4paper]{revtex4}

\usepackage[dvips]{graphicx}
\usepackage{dcolumn}
\usepackage{bm}
\usepackage{epsfig}
\usepackage{amsfonts}
\usepackage{amssymb,amscd}

\def\lsim{\raise0.3ex\hbox{$<$\kern-0.75em\raise-1.1ex\hbox{$\sim$}}}
\def\gsim{\raise0.3ex\hbox{$>$\kern-0.75em\raise-1.1ex\hbox{$\sim$}}}

\def\beq{\begin{equation}}
\def\eeq{\end{equation}}
\def\bea{\begin{eqnarray}}
\def\eea{\end{eqnarray}}
\def\bq{\begin{quote}}
\def\eq{\end{quote}}

\newcommand{\rr}{\mbox{\boldmath $r$}}

\newcommand{\rb}{\mbox{\boldmath $b$}}

\newcommand{\rd}{\mbox{\boldmath $\Delta$}}

\def\gappeq{\mathrel{\rlap {\raise.5ex\hbox{$>$}}
{\lower.5ex\hbox{$\sim$}}}}

\def\lappeq{\mathrel{\rlap{\raise.5ex\hbox{$<$}}
{\lower.5ex\hbox{$\sim$}}}}

\def\Toprel#1\over#2{\mathrel{\mathop{#2}\limits^{#1}}}

\begin{document}

%\preprint{Version 1.2}

\title{Investigation of diffractive photoproduction of $J/\Psi$ in hadronic collisions}

\author{V.~P. Gon\c{c}alves $^1$, B. D. Moreira $^2$ and F. S. Navarra $^2$}
%\email{barros@ufpel.edu.br}
\affiliation{$^1$ High and Medium Energy Group, \\
Instituto de F\'{\i}sica e Matem\'atica, Universidade Federal de Pelotas\\
Caixa Postal 354, CEP 96010-900, Pelotas, RS, Brazil \\ $^2$ Instituto de F\'{\i}sica, Universidade de S\~{a}o Paulo,
C.P. 66318,  05315-970 S\~{a}o Paulo, SP, Brazil}

\date{\today}

\begin{abstract}

In this work we study the diffractive photoproduction of $J/\Psi$  in proton-proton, proton - nucleus and nucleus - nucleus  collisions at LHC energies using the color dipole formalism and different models for the forward dipole - target scattering amplitude. Our goal is to estimate the theoretical uncertainty present in the current predictions  in the literature. Our results are compared with the experimental data and predictions for higher energies are presented.
 
\end{abstract}

\maketitle

%%%%%%%%%%%%%%%%%%%%%%%%%%%%%%%%%%%%%%%%%%%%%%%%%%%%%%%%%%%%%%%%%%%%%%%%%%%%%%%%
\section{Introduction}
\label{intro}
Since the pioneering studies \cite{klein_prc,gluon,strikman} on diffractive vector meson production in ultra peripheral heavy ion collisions (UPHIC) about fourteen years ago, a large number of papers on the subject has been published considering  several improvements in the theoretical description \cite{outros_klein,vicmag_mesons1,outros_vicmag_mesons,outros_frankfurt,Schafer,vicmag_update,gluon2,motyka_watt,Lappi,griep,Guzey,Martin,cisek}  and experimental analysis \cite{cdf,star,phenix,alice,alice2,lhcb}. Recent experimental results from CDF  \cite{cdf} at Tevatron, STAR \cite{star} and PHENIX \cite{phenix} at RHIC and ALICE \cite{alice,alice2} and LHCb \cite{lhcb,lhcb2} at LHC have demonstrated that the study of photon-hadron interactions in these colliders is feasible and that the data can be used to constrain the description of the hadronic structure at high energies (For reviews see Ref. \cite{upc}).  In particular, the recent theoretical studies performed  in Refs. \cite{Guzey,Martin} have demonstrated that the description of the ALICE and LHCb data using the collinear factorization formalism is strongly dependent on the choice of the parameterization of the gluon distribution and on the magnitude of the nuclear shadowing effects (For a similar analysis using the $k_T$-factorization approach see Ref. \cite{cisek}). On the other hand, as  originally proposed in Ref. \cite{vicmag_mesons1}, the diffractive  photoproduction of vector mesons in hadronic collisions  can also be studied in the color dipole formalism \cite{nik}  and used to constrain the magnitude of the nonlinear effects which are predicted to be present at high parton densities by the QCD dynamics at high energies. In this formalism the predictions for the vector meson production in hadronic collisions are strongly dependent on the model used to calculate the forward dipole - target scattering amplitude $\cal{N}$. As the current color dipole predictions for the diffractive photoproduction of $J/\Psi$ in hadronic collisions \cite{outros_vicmag_mesons,vicmag_update,motyka_watt,Lappi,griep} consider distinct models for this quantity, photon spectrum, skewness corrections and/or for the vector meson wavefunction, a direct comparison between its predictions is a hard task. Our goal in this paper is to compare the predictions obtained considering different models for $\cal{N}$ assuming a unified treatment for the photon spectrum and $J/\Psi$ wavefunction.   
It allow us to estimate the theoretical uncertainty present in the current predictions in the literature. After  showing  that the different models used in our calculations are able to describe the $\gamma p$ HERA data, we  present predictions for the photonuclear $J/\Psi$  production which could be probed in future electron - ion colliders \cite{eA}. Moreover, we compare our predictions for the diffractive 
photoproduction of $J/\Psi$ in $pp$ and $PbPb$ collisions  with recent LHCb and ALICE data, respectively. Finally, we present predictions for future LHC energies and $pPb$ collisions.

The paper is organized as follows. In the next section we present a brief review of photon - hadron 
interactions in $pp$ and $PbPb$ collisions, as well as of the diffractive photoproduction of $J/\Psi$ in the color dipole formalism. We also present the models for the dipole - target scattering 
amplitude used in our calculations. In Section \ref{resultados} we present our predictions 
for the diffractive photoproduction of $J/\Psi$ in $pp/pPb/PbPb$ collisions and a comparison 
with the ALICE and LHCb data is also  shown. Finally, in Section \ref{sumario} we summarize our main conclusions.

\section{Formalism}
\label{formulas}

Initially let us present a brief review of photon - hadron interactions in hadronic collisions.
Let us consider a hadron-hadron interaction at large impact parameter 
($b > R_{h_1} + R_{h_2}$) and at ultra relativistic energies. In this regime we expect the 
electromagnetic interaction to be dominant.
In  heavy ion colliders, the heavy nuclei give rise to strong electromagnetic fields due to 
the coherent action of all protons in the nucleus. In a similar way, this kind of interaction 
also occurs in ultra relativistic  protons in $pp(\bar{p})$ colliders.
The photon stemming from the electromagnetic field
of one of the two colliding hadrons can interact  directly with the other hadron 
(photon-hadron process).  The total
cross section for a given process can be factorized in terms of the equivalent flux of 
photons of the hadron projectile and  the  photon-target production cross 
section \cite{upc}.  The cross section for  diffractive $J/\Psi$ photoproduction  in a   
hadron-hadron collision is  given by,
\begin{eqnarray}
\sigma (h_1 h_2 \rightarrow h_1 \otimes J/\Psi \otimes h_2) = 
\int \limits_{\omega_{min}}^{\infty} d\omega  \,
\frac{dN_{\gamma/h_1}(\omega)}{d\omega}\,
\sigma_{\gamma h_2 \rightarrow J/\Psi h_2} (W_{\gamma h_2}^2)  \nonumber \\ +
 \int \limits_{\omega_{min}}^{\infty} d\omega  \,
\frac{dN_{\gamma/h_2}(\omega)}{d\omega}\,
\sigma_{\gamma h_1 \rightarrow J/\Psi h_1} (W_{\gamma h_1}^2),
\label{sighh}
\end{eqnarray}
where $\otimes$ represents a rapidity gap in the final state, $\omega$ is the photon 
energy in the collider frame,   $\omega_{min}=M_{J/\Psi}^2/4\gamma_L m_p$,  $\gamma_L$ is the Lorentz boost  of a 
single beam, $\frac{dN_{\gamma}}{d\omega}$ is the equivalent photon flux, 
$W_{\gamma h}^2=2\,\omega \sqrt{s_{\mathrm{NN}}}$  and $\sqrt{s_{\mathrm{NN}}}$ is 
 the c.m.s energy of the
hadron-hadron system.
Considering the requirement that  photoproduction
is not accompanied by hadronic interaction (ultra-peripheral
collision) an analytic approximation for the equivalent photon flux of a nuclei can be 
calculated and  given by \cite{upc}
\begin{eqnarray}
\frac{dN_{\gamma/A}\,(\omega)}{d\omega}= \frac{2\,Z^2\alpha_{em}}{\pi\,\omega}\, \left[\bar{\eta}\,K_0\,(\bar{\eta})\, K_1\,(\bar{\eta})+ \frac{\bar{\eta}^2}{2}\,{\cal{U}}(\bar{\eta}) \right]\,
\label{fluxint}
\end{eqnarray}
where   $\bar{\eta}=\omega\,(R_{h_1} + R_{h_2})/\gamma_L$ and  
${\cal{U}}(\bar{\eta}) = K_1^2\,(\bar{\eta})-  K_0^2\,(\bar{\eta})$. 
Eq. (\ref{fluxint}) will be used in our calculations of the diffractive   photoproduction of $J/\Psi$ in $pPb$ and $PbPb$ 
collisions. On the other hand, for   proton-proton collisions, we assume that the  photon 
spectrum of a relativistic proton is given by  \cite{Dress},
\begin{eqnarray}
\frac{dN_{\gamma/p}(\omega)}{d\omega} =  \frac{\alpha_{\mathrm{em}}}{2 \pi\, \omega} \left[ 1 + \left(1 -
\frac{2\,\omega}{\sqrt{s_{NN}}}\right)^2 \right] 
\left( \ln{\Omega} - \frac{11}{6} + \frac{3}{\Omega}  - \frac{3}{2 \,\Omega^2} + \frac{1}{3 \,\Omega^3} \right) \,,
\label{eq:photon_spectrum}
\end{eqnarray}
with the notation $\Omega = 1 + [\,(0.71 \,\mathrm{GeV}^2)/Q_{\mathrm{min}}^2\,]$ and 
$Q_{\mathrm{min}}^2= \omega^2/[\,\gamma_L^2 \,(1-2\,\omega /\sqrt{s_{NN}})\,] \approx 
(\omega/
\gamma_L)^2$.

The main input in our calculations is the diffractive photoproduction cross section $\sigma_{\gamma h \rightarrow J/\Psi h}$. 
In what follows we describe the $\gamma h$  scattering in the dipole frame, in which most of the energy 
is
carried by the hadron, while the  photon  has
just enough energy to dissociate into a quark-antiquark pair
before the scattering. In this representation the probing
projectile fluctuates into a
quark-antiquark pair (a dipole) with transverse separation
$\rr$ long before the interaction, which then
scatters off the hadron \cite{nik}. Initially, let us consider a photon - proton interaction ($h = p$). 
In the dipole picture the   amplitude for the diffractive photoproduction of an exclusive final state, 
such as a $J/\Psi$, in a $\gamma p$ collision is given by 
(See e.g. Refs. \cite{nik,vicmag_mesons,KMW,armesto_amir})
\begin{eqnarray}
 {\cal A}^{\gamma p \rightarrow J/\Psi p}(x,\Delta)  =  i
\int dz \, d^2\rr \, d^2\rb  e^{-i[\rb-(1-z)\rr].\rd} 
 \,\, (\Psi_{J/\Psi}^* \Psi) \,\,2 {\cal{N}}_p(x,\rr,\rb)
\label{sigmatot2}
\end{eqnarray}
where $(\Psi_{J/\Psi}^* \Psi)$ denotes the overlap of the photon and $J/\Psi$   
transverse wave functions. The variable  $z$ $(1-z)$ is the
longitudinal momentum fractions of the quark (antiquark),  $\Delta$ denotes the transverse 
momentum lost by the outgoing proton ($t = - \Delta^2$) and $x$ is the Bjorken variable. 
The variable $\rb$ is the transverse distance from the center of the target to the center of mass of the 
$q \bar{q}$  dipole and the factor  in the exponential  arises when one takes into account 
non-forward corrections to the wave functions \cite{non}.
 Moreover, ${\cal{N}}_p (x,\rr,\rb)$ denotes the forward  amplitude of  the scattering of a dipole of size $\rr$ on the proton, which is  directly related to  the QCD 
dynamics (see below). The total cross section  for  diffractive photoproduction of $J/\Psi$ is given by
\begin{eqnarray}
\sigma (\gamma p \rightarrow J/\Psi p) = \frac{1}{16\pi}  \int dt \,  |{\cal{A}}^{\gamma p \rightarrow J/\Psi p}(x,\Delta)|^2\,R_g^2(1 + \beta^2)\,\,
\label{totalcs}
\end{eqnarray}
where $\beta$ is the ratio of real to imaginary parts of the scattering
amplitude,  which can be obtained  using dispersion relations
 $Re {\cal A}/Im {\cal A}=\mathrm{tan}\,(\pi \lambda_e/2)$. Moreover, $R_g$ is the skewness factor, which is associated to the fact that the gluons attached to the $q\bar{q}$ pair can carry different light-cone fractions $x$, $x^{\prime}$ of the proton. In the limit that $x^{\prime} \ll x \ll 1$ and at small $t$ and assuming that the gluon density has a power-law form, it is given by \cite{Shuvaev:1999ce}
\begin{eqnarray}
\label{eq:Rg}
  R_g(\lambda_e) = \frac{2^{2\lambda_e+3}}{\sqrt{\pi}}
\frac{\Gamma(\lambda_e+5/2)}{\Gamma(\lambda_e+4)}, 
  \quad\text{with} \quad \lambda_e \equiv 
\frac{\partial\ln\left[\mathcal{A}(x,\,\Delta)\right]}{\partial\ln(1/x)}.
\end{eqnarray}
As demonstrated in Ref. \cite{harland} the estimate obtained using this approximation for $R_g$ is strongly dependent on the parton distribution used in the calculation. Furthermore, the incorporation of the skewness correction at small-$x$ in the dipole models still is an open question. Consequently, the factor $R_g$ as given in Eq. (\ref{eq:Rg}) should be considered a phenomenological estimate.

 If we assume that the  nucleus 
scatters elastically (coherent production) and also that  the scattering happens  
in the high energy regime (large coherence length: $l_c \gg R_A$) the total photon-nucleus cross 
section is given by:  
\cite{vmprc,kop1}
\begin{eqnarray}
\sigma^{coh}\, (\gamma A \rightarrow J/\Psi A)  =  \int d^2\rb \left[
\int d^2\rr
 \int dz (\Psi_{J/\Psi}^* \Psi) \, \mathcal{N}_A(x,\rr,\rb)\right]^2
\label{totalcscoe}
\end{eqnarray}
where  $ {\cal N}_A (x, \rr, \rb)$ is the forward dipole-nucleus scattering amplitude.
 Finally, the photon wave functions appearing in Eq. (\ref{sigmatot2}) are well known in literature 
\cite{KMW}. For the meson wave function, we have considered the Gauss-LC  model \cite{KMW} 
which is a simplification of the DGKP wave functions \cite{dgkp}. The motivation for this choice is its 
simplicity and the fact that, for the present purposes, the final results are not very sensitive 
these details. In  photoproduction, this leads only to an  uncertainty  of a few percent in the 
overall normalization.  The parameters of the meson wave function can be found in Ref. \cite{KMW}.

The scattering amplitude ${\cal{N}}(x,\rr,\rb)$   contains all
information about the target and the strong interaction physics.
In the Color Glass Condensate (CGC)  formalism \cite{CGC,BAL}, it  encodes all the
information about the
non-linear and quantum effects in the hadron wave function. It can be obtained by solving 
an appropriate evolution
equation in the rapidity $Y\equiv \ln (1/x)$, which in its  simplest form is the 
Balitsky-Kovchegov (BK) equation \cite{BAL,kov}. In recent years,  the running coupling corrections to BK evolution kernel were explicitly calculated  \cite{kovwei1,balnlo},  including the  $\alpha_sN_f$ corrections to 
the kernel to all orders, and its solution studied in detail  \cite{javier_kov,javier_prl}. Basically, one has that the running of the coupling reduces the speed of the evolution to values compatible with experimental $ep$ HERA data \cite{bkrunning,weigert}.   In Ref. \cite{bkrunning} the translational invariance approximation was assumed, which implies ${\cal{N}}(x,\rr,\rb) = {\cal{N}}(x,\rr) S(\rb)$, with the normalization of the dipole cross section being fitted to data and two distinct initial conditions, inspired in the Golec Biernat-Wusthoff (GBW) \cite{GBW} and McLerran-Venugopalan (MV) \cite{MV} models, were considered. The predictions resulted to be almost independent of the initial conditions and, besides, it was observed that it is impossible to describe the experimental data using only the linear limit of the BK equation, which is equivalent to the Balitsky-Fadin-Kuraev-Lipatov (BFKL) equation \cite{bfkl}. In what follows we use the solution obtained in Ref. \cite{bkrunning} considering the MV initial condition, denoting the corresponding predictions by rcBK hereafter. It is important to emphasize that although a complete analytical solution of the BK equation is still lacking, its main properties are known: (a) for the interaction of a small dipole ($|\rr| \ll
1/Q_s$), ${\cal{N}}(x,\rr,\rb) \approx \rr^2$, implying  that
this system is weakly interacting; (b) for a large dipole ($|\rr| \gg
1/Q_s$), the system is strongly absorbed and therefore
${\cal{N}}(x,\rr,\rb) \approx 1$. The typical momentum scale, $Q_s^2\propto x^{-\lambda}\,(\lambda\approx 0.3)$, is the so called saturation scale. This property is associated  to the
large density of saturated gluons in the hadron wave function. In the last years, several groups have constructed phenomenological models which satisfy the asymptotic behavior of the BK equation in order to fit the HERA and RHIC data (See e.g. Refs. \cite{GBW,dipolos10,iim,kkt,dhj,Goncalves:2006yt,buw,KMW}). 
For comparison, in what follows we will also  use the GBW model \cite{GBW}, which  assumes that 
${\cal{N}}_p(x,\rr,\rb) = {\cal{N}}_p(x,\rr) S(\rb)$ with ${\cal{N}}_p(x,\rr)=1-e^{-\rr^2Q_{s,p}^2(Y)/4}$ and $Q_{s,p}^2(Y)=\left(x_0/x\right)^{\lambda}$, with the parameters $x_0$ and $\lambda$ determined by the fit to the HERA data. Moreover, we also consider the 
b-CGC model proposed in Ref. \cite{KMW}, which improves the Iancu - Itakura - Munier (IIM) model 
 \cite{iim} with  the inclusion of   the impact parameter dependence in the dipole - proton scattering amplitude.   Following \cite{KMW} we have:
\begin{eqnarray}
\mathcal{N}_p(x,\rr,{\rb}) =   
\left\{ \begin{array}{ll} 
{\mathcal N}_0\, \left(\frac{ r \, Q_{s,p}}{2}\right)^{2\left(\gamma_s + 
\frac{\ln (2/r Q_{s,p})}{\kappa \,\lambda \,Y}\right)}  & \mbox{$r Q_{s,p} \le 2$} \\
 1 - \exp^{-A\,\ln^2\,(B \, r \, Q_{s,p})}   & \mbox{$r Q_{s,p}  > 2$} 
\end{array} \right.
\label{eq:bcgc}
\end{eqnarray}
with  $Y=\ln(1/x)$ and $\kappa = \chi''(\gamma_s)/\chi'(\gamma_s)$, where $\chi$ is the 
LO BFKL characteristic function.  The coefficients $A$ and $B$  
are determined uniquely from the condition that $\mathcal{N}_p(x,\rr,\rb)$, and its derivative 
with respect to $rQ_s$, are continuous at $rQ_s=2$. 
In this model, the proton saturation scale $Q_{s,p}$ depends on the impact parameter:
\begin{equation} 
  Q_{s,p}\equiv Q_{s,p}(x,{\rb})=\left(\frac{x_0}{x}\right)^{\frac{\lambda}{2}}\;
\left[\exp\left(-\frac{{b}^2}{2B_{\rm CGC}}\right)\right]^{\frac{1}{2\gamma_s}}.
\label{newqs}
\end{equation}
The parameter $B_{\rm CGC}$  was  adjusted to give a good 
description of the $t$-dependence of exclusive $J/\psi$ photoproduction.  
The factors $\mathcal{N}_0$ and  $\gamma_s$  were  taken  to be free. In this 
way a very good description of  $F_2$ data was obtained. 
One of the parameter set  which is going to be used here is the one presented in the second line of
Table II of  \cite{watt}:  $\gamma_s = 0.46$, $B_{CGC} = 7.5$ GeV$^{-2}$,
$\mathcal{N}_0 = 0.558$, $x_0 = 1.84 \times 10^{-6}$ and $\lambda = 0.119$.
More recently, the parameters of this model have been updated in Ref. \cite{amir} (considering the 
recently released high precision combined HERA data), becoming  $\gamma_s = 0.6599$, $B_{CGC} = 5.5$ GeV$^{-2}$,
$\mathcal{N}_0 = 0.3358$, $x_0 = 0.00105 \times 10^{-5}$ and $\lambda = 0.2063$.
In what follows we will use these two sets of parameters in our calculations, with the resulting predictions being denoted bCGC and bCGC NEW, respectively.

In order to estimate the diffractive photoproduction of $J/\Psi$  in $PbPb$ collisions we need to specify the 
forward dipole - nucleus scattering amplitude, $\mathcal{N}_A(x,\rr,\rb)$.  
Following \cite{vmprc} we will use in our calculations  the model proposed in Ref. 
\cite{armesto}, which describes  the current  experimental data on the nuclear 
structure function as well as includes the  impact parameter dependence in the dipole 
nucleus cross section. In this model the forward dipole-nucleus amplitude is given by
\begin{eqnarray}
{\cal{N}}_A(x,\rr,\rb) = 1 - \exp \left[-\frac{1}{2}  \, \sigma_{dp}(x,\rr^2) 
\,T_A(\rb)\right] \,\,,
\label{enenuc}
\end{eqnarray}
where $\sigma_{dp}$ is the dipole-proton cross section given by
\begin{eqnarray}
\sigma_{dp} (x,\rr^2) = 2 \int d^2\rb \,\,\mathcal{N}_p(x,\rr,{\rb})  
\end{eqnarray}
 and $T_A(\rb)$ is the nuclear profile 
function, which is obtained from a 3-parameter Fermi distribution for the nuclear
density normalized to $A$.
The above equation
%, based on the Glauber-Gribov formalism \cite{gribov},  
sums up all the 
multiple elastic rescattering diagrams of the $q \overline{q}$ pair
and is justified for large coherence length, where the transverse separation $\rr$ of 
partons 
in the multiparton Fock state of the photon becomes a conserved quantity, {\it i.e.} 
the size 
of the pair $\rr$ becomes eigenvalue
of the scattering matrix.  In what follows we will compute  $\mathcal{N}_A$ considering 
the different models for the dipole - proton scattering amplitude discussed before.

\begin{figure}
\begin{tabular}{cc}
\includegraphics[scale=0.35]{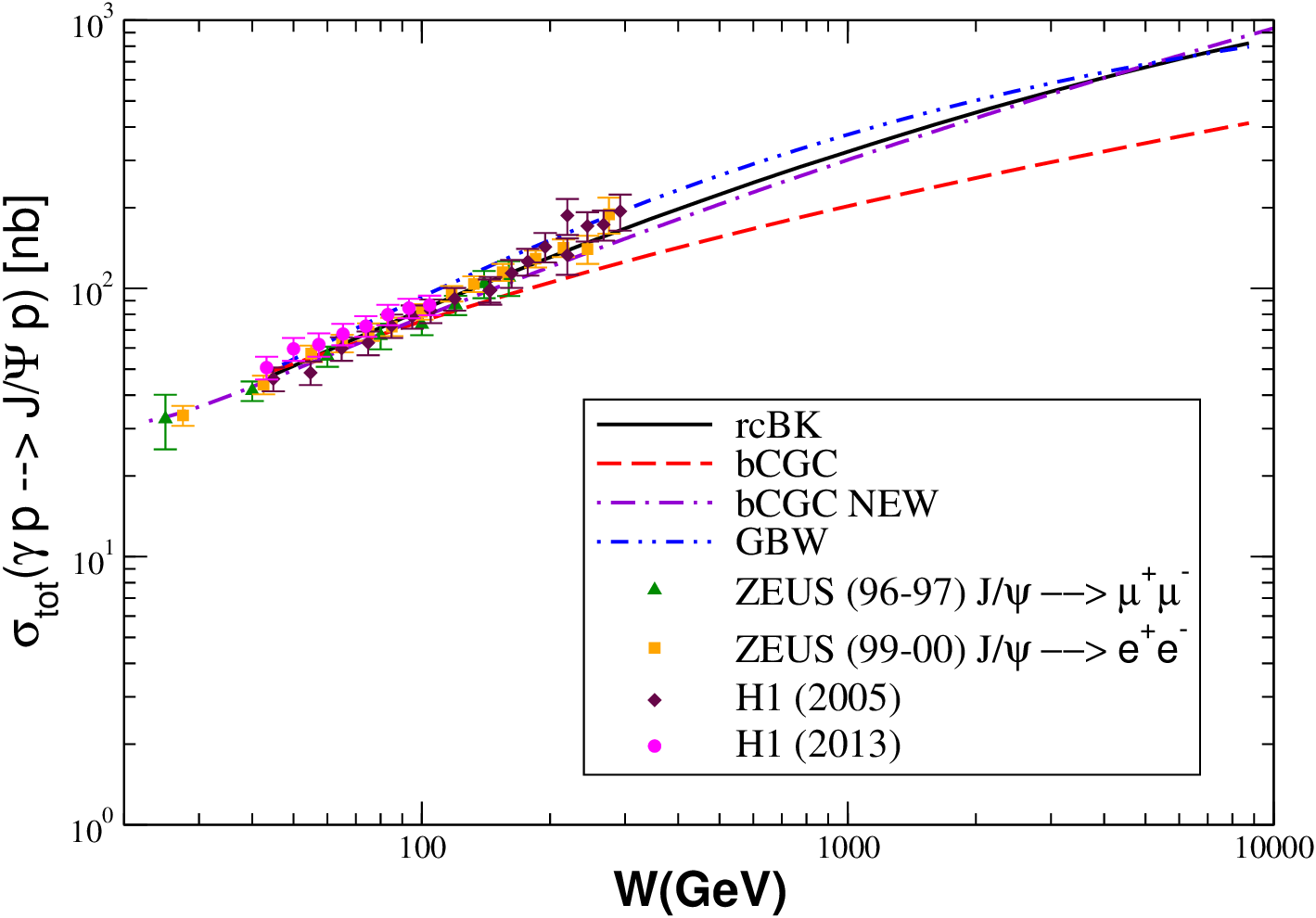} & 
\includegraphics[scale=0.35]{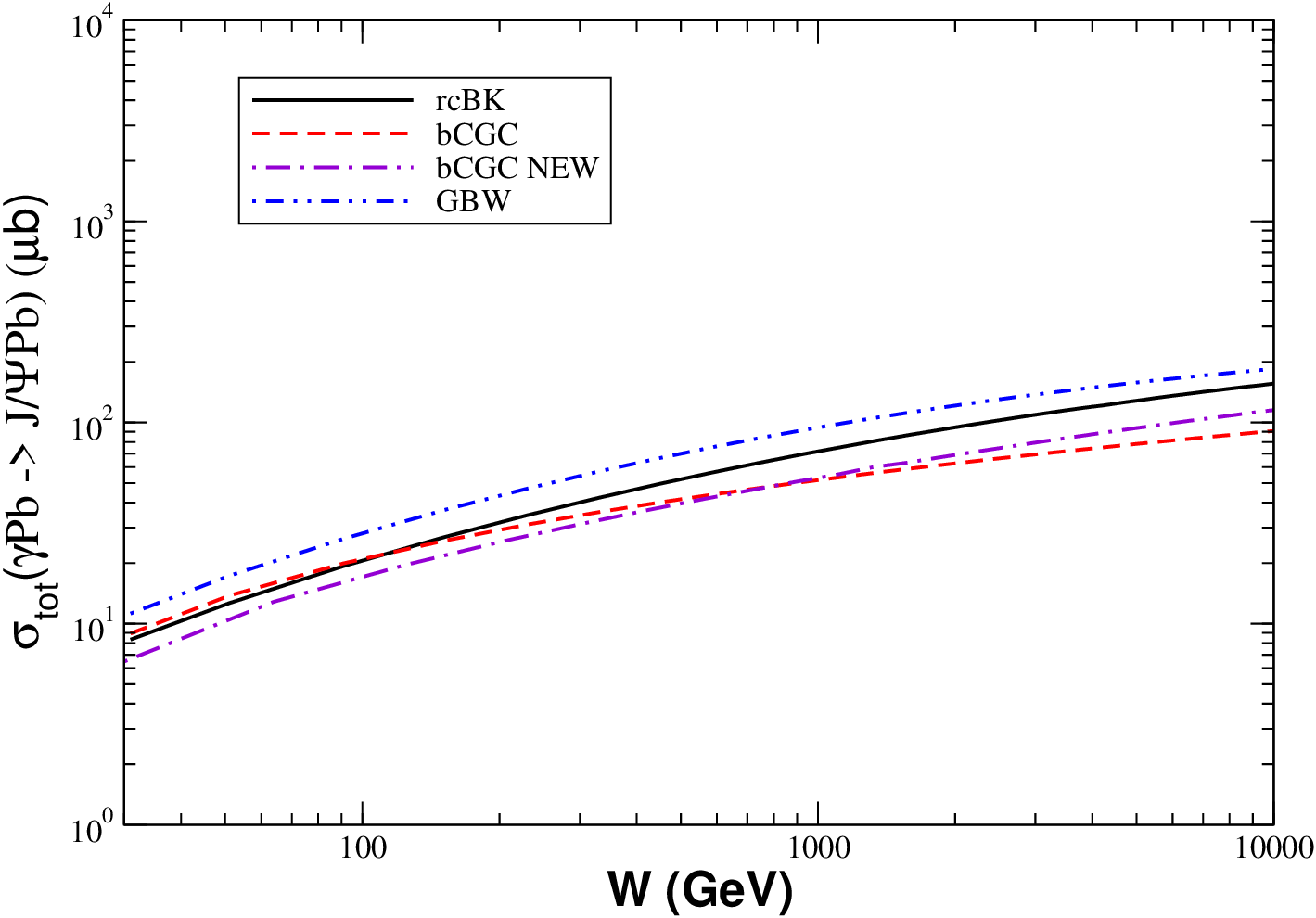}
\end{tabular}
\caption{(Color online) Diffractive photoproduction of $J/\Psi$ in (a)  $\gamma p$  
and (b) $\gamma Pb$ collisions. Data from HERA \cite{hera_data}. }
\label{fig1}
\end{figure}

\section{Results}
\label{resultados}

In Fig. \ref{fig1} we present our predictions for the diffractive photoproduction of $J/\Psi$ in (a) $\gamma p$  
and (b) $\gamma Pb$  collisions considering as input the different models for the dipole - proton scattering amplitude discussed in the previous section. For the $\gamma p$ case , we 
find  that the distinct models are able to describe the low energy data, but predict different behaviours at high energies. It is important to emphasize that the parameterization for the solution of the BK equation proposed in Ref. \cite{bkrunning} is only valid for $x \le 10^{-2}$.  This restricts its use for $W_{\gamma p} \ge 30$ GeV. Moreover, we find that the improved version of the bCGC (bCGC NEW) predicts a larger growth for the total cross section in comparison with the previous one, which becomes its predictions compatible with the HERA data.  For $\gamma Pb$ collisions, we find that the rcBK and bCGC predictions are similar at low energies, but differ by a factor 1.8 at larger energies. On the other hand, the bCGC NEW prediction can be considered a lower bound for the diffractive photoproduction of $J/\Psi$ at low energies, while the GBW one is an upper bound. As already pointed in Ref. \cite{vmprc},  future measurements in an $eA$ collider \cite{eA} could discriminate between these models.

Let us now estimate the rapidity distribution in  diffractive photoproduction 
of $J/\Psi$ in hadronic collisions. 
The rapidity ($Y$) distribution of the $J/\Psi$ in the final state can be directly computed from Eq. (\ref{sighh}), by using its  relation with the photon energy $\omega$, i.e. $Y\propto \ln \, ( \omega/m_{J/\Psi})$.  Explicitly, the rapidity distribution is written down as,
\begin{eqnarray}
\frac{d\sigma \,\left[h_1 + h_2 \rightarrow   h_1 \otimes J/\Psi \otimes h_2\right]}{dY} = \left[\omega \frac{dN}{d\omega}|_{h_1}\,\sigma_{\gamma h_2 \rightarrow J/\Psi \otimes h_2}\left(\omega \right)\right]_{\omega_L} + \left[\omega \frac{dN}{d\omega}|_{h_2}\,\sigma_{\gamma h_1 \rightarrow J/\Psi \otimes h_1}\left(\omega \right)\right]_{\omega_R}\,
\label{dsigdy}
\end{eqnarray}
where $\otimes$ represents the presence of a rapidity gap in the final state and $\omega_L \, (\propto e^{-Y})$ and $\omega_R \, (\propto e^{Y})$ denote photons from the $h_1$ and $h_2$ hadrons, respectively.  As the photon fluxes, Eqs. (\ref{fluxint}) and (\ref{eq:photon_spectrum}), have support at small values of $\omega$, decreasing exponentially at large $\omega$, the first term on the right-hand side of the Eq. (\ref{dsigdy}) peaks at positive rapidities while the second term peaks at negative rapidities. Consequently, given the photon flux, the study of the rapidity distribution can be used to constrain  the photoproduction cross section  at  a given energy. Moreover, in contrast to the total rapidity distributions for $pp$ and $PbPb$ collisions, which will be symmetric about midrapidity ($Y=0$), $d\sigma/dY$ will be asymmetric in $pPb$ collisions due to the differences between the fluxes and process cross sections.
It is important to emphasize that in our calculations we will disregard   soft interactions which lead to an extra production of particles that destroy the rapidity gap in the final state. 
The inclusion of these additional absorption effects can be parametrized in terms of a multiplicative factor denoted rapidity gap survival probability, $S^2$, which corresponds to the probability of the scattered proton not to dissociate due to the secondary interactions. 
In Ref. \cite{Schafer}  the authors have estimated $S^2$ and obtained that in $pp/p\bar{p}$ collisions it is $ \sim 0.8 - 0.9$,  depending  on the  rapidity of the vector meson (See also Refs. \cite{Guzey,Martin}). Moreover, in the particular case of nuclear interactions, we also disregard a possible reduction of the  nuclear photon flux associated to the diffusion edge of the nucleus.

\begin{figure}
\begin{tabular}{cc}
\includegraphics[scale=0.35]{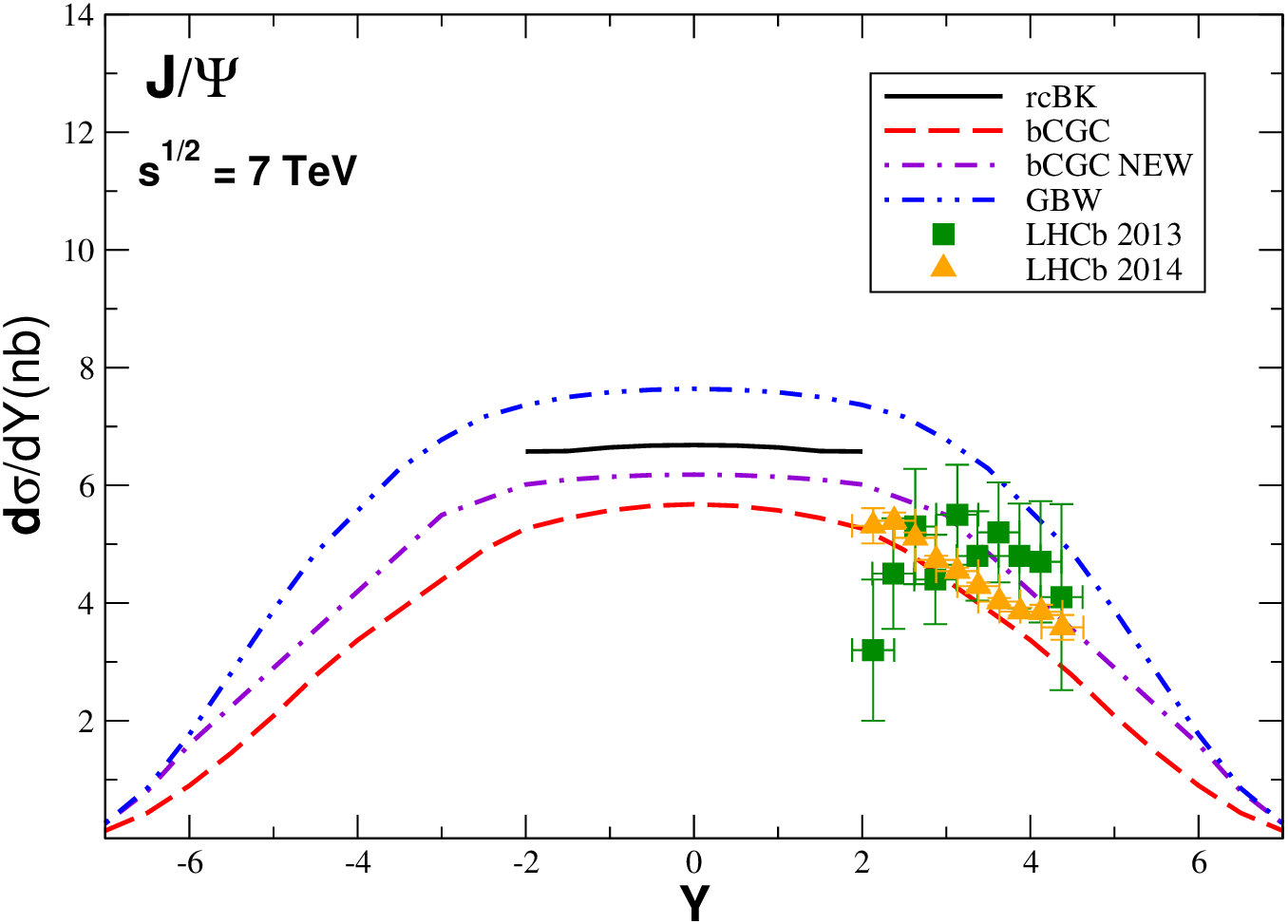} & 
\includegraphics[scale=0.35]{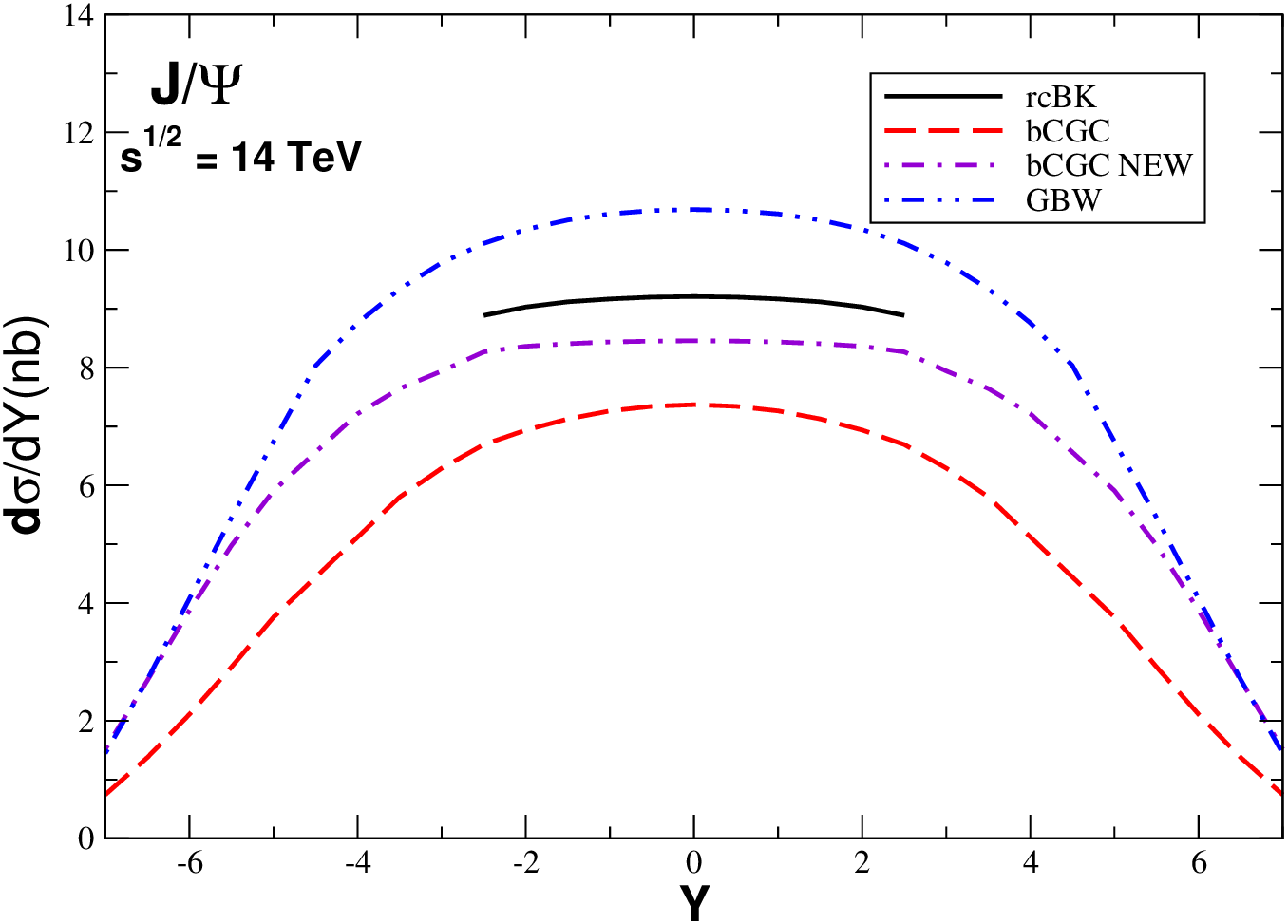}
\end{tabular}
\caption{(Color online) Rapidity distribution for the diffractive photoproduction of  $J/\Psi$  in $pp$ collisions  at (a) $\sqrt{s}=7$ TeV  and  (b) $\sqrt{s}=14$ TeV. Data from LHCb Collaboration \cite{lhcb,lhcb2}.}
\label{fig2}
\end{figure}

In Fig. \ref{fig2} we present our predictions for the rapidity distribution for the diffractive photoproduction of  $J/\Psi$  in $pp$ collisions  at (a) $\sqrt{s}=7$ TeV and  (b)  $\sqrt{s}=14$ TeV. 
Due to the limitation in the $x$-range of the rcBK solution, we  are only able to present its predictions for a restricted rapidity range. We obtain that the differences between the predictions observed in Fig. \ref{fig1} are also presented in the rapidity distribution, with the GBW (bCGC) prediction being an upper (lower) bound for the predictions at $Y = 0$. In particular, the predictions differ by $\approx 30$ \% for central rapidities  at $\sqrt{s} = 7$ TeV. For the rapidity range probed by the LHCb Collaboration the difference is larger ($\approx 50$ \%). We obtain that the bCGC and bCGC NEW predictions agree  with the data from LHCb Collaboration \cite{lhcb,lhcb2}. As demonstrated in Fig. \ref{fig2} (b), these differences increase with the energy. This motivates future experimental analysis of this process in order to constrain the dipole - proton scattering amplitude and, consequently, the QCD dynamics at high energies.

\begin{figure}
\begin{tabular}{cc}
\includegraphics[scale=0.35]{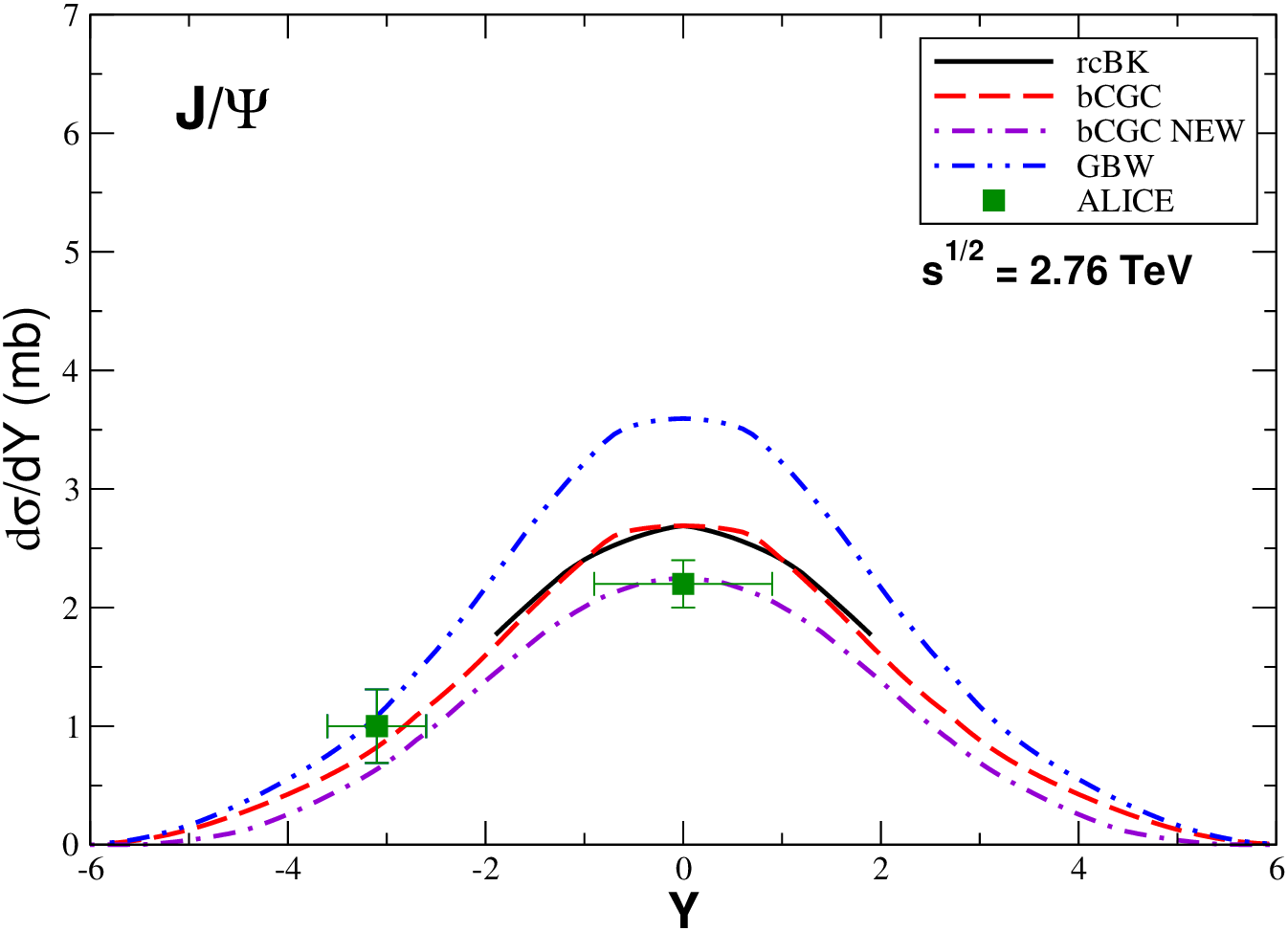} & 
\includegraphics[scale=0.35]{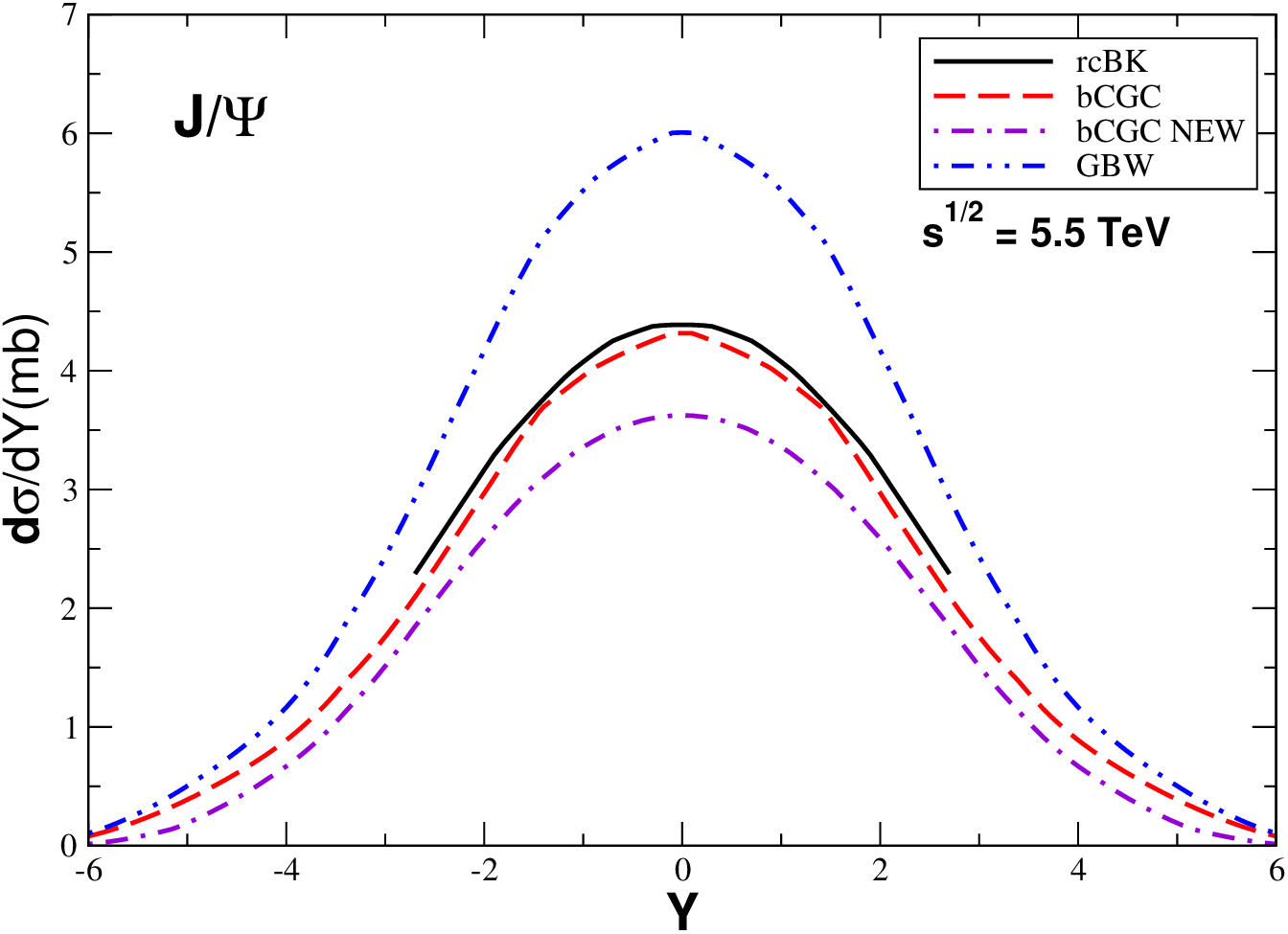}
\end{tabular}
\caption{(Color online) Rapidity distribution for the diffractive photoproduction of  $J/\Psi$  in $PbPb$ collisions  at (a) $\sqrt{s}=2.76$ TeV and   (b) $\sqrt{s}=5.5$ TeV. Data from ALICE Collaboration \cite{alice,alice2}.}
\label{fig3}
\end{figure}

\begin{figure}
%\begin{tabular}{ccc}
\includegraphics[scale=0.35]{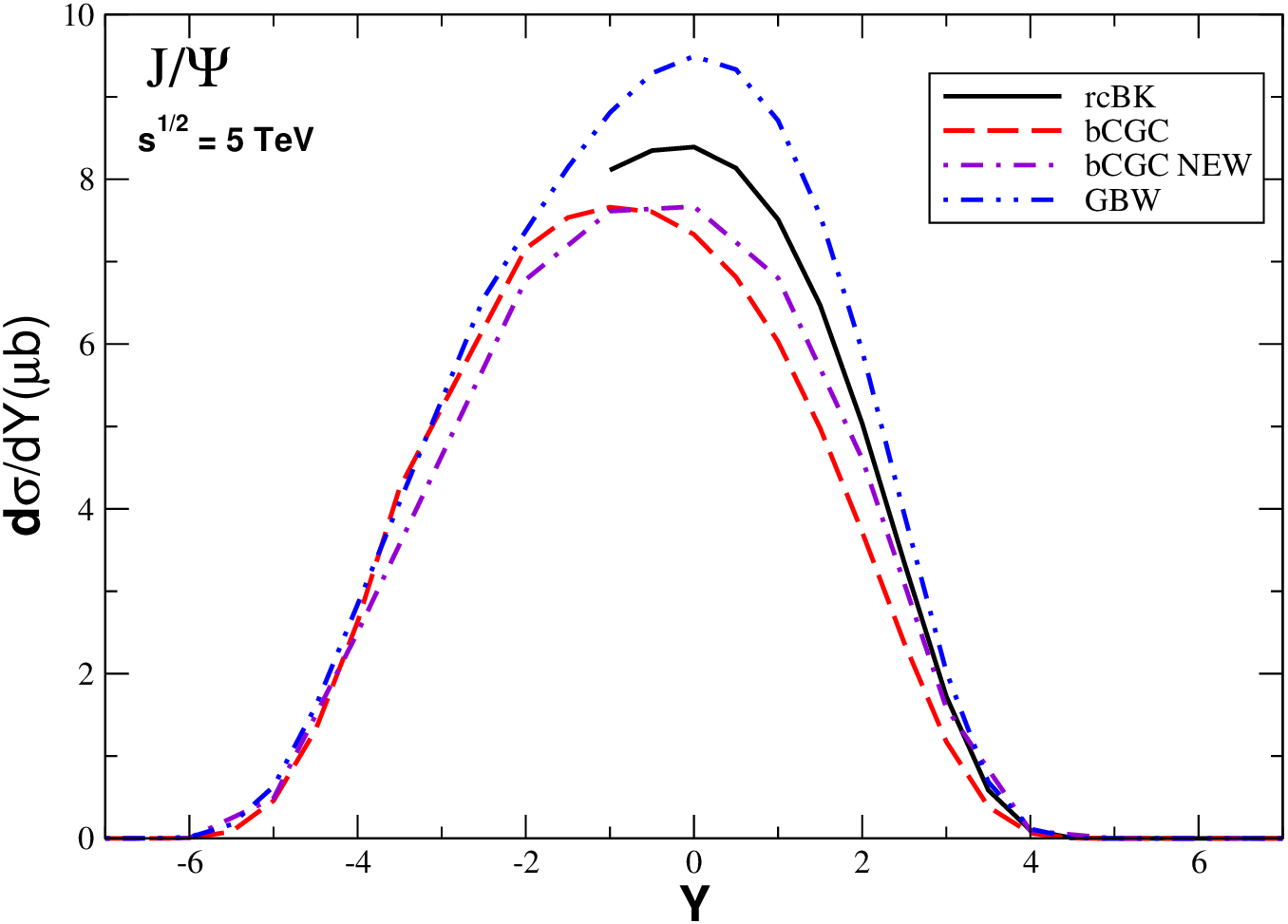}
\caption{(Color online) Rapidity distribution for the diffractive photoproduction of  $J/\Psi$  in $pPb$ collisions  at $\sqrt{s}=5$ TeV. }
\label{fig4}
\end{figure}

In Fig. \ref{fig3} we present our predictions for the rapidity distribution for the diffractive photoproduction of  $J/\Psi$  in $PbPb$ collisions  at (a) $\sqrt{s}=2.76$ TeV  and (b)  $\sqrt{s}=5.5$ TeV. In this case the cross sections are calculated in terms of the dipole - nucleus scattering amplitude given in Eq. (\ref{enenuc}).  Similarly to the $pp$ case, we obtain that the distinct predictions largely differ at central rapidities, which is directly associated to the behavior observed in Fig. \ref{fig1} (b) for $\gamma Pb$ collisions. We obtain that the bCGC NEW prediction is able to describe the current  ALICE data \cite{alice,alice2}, in contrast with the other predictions which overestimate the data for $Y = 0$.  In particular, the rcBK prediction is not able to describe the data, in agreement with the results obtained in Ref. \cite{Lappi}. As observed in Fig. \ref{fig3} (b) the difference between the predictions is amplified at larger energies.  
In Fig. \ref{fig4} we present our predictions for the rapidity distribution for the diffractive photoproduction of  $J/\Psi$  in $pPb$ collisions  at $\sqrt{s}=5$ TeV. As expected, the rapidity distribution is asymmetric about midrapidity ($Y=0$), being dominated by $\gamma p$ interactions, due to the $Z^2$ enhancement present in the nuclear photon spectrum. We observe that the predictions differ by $\approx 35$ \% at $Y= 0$. Finally, in Table \ref{tab} we present our predictions for the total cross section for the diffractive photoproduction of $J/\Psi$  in $pp$, $pPb$ and $PbPb$  collisions at  LHC energies. As expected from our analysis of the rapidity distributions, the predictions for the total cross sections are largely distinct.

\section{Summary}
\label{sumario}

The recent experimental data from RHIC, Tevatron and LHC have demonstrated that the study of photon - hadron interactions in hadron - hadron collisions in order to constrain the QCD dynamics at high energies is feasible. They have motivated the proposition of new observables which can be studied in these processes and the  improvement its theoretical description. In particular, in the last year, several studies have been performed considering the collinear formalism and the DGLAP evolution, which demonstrated that the diffractive photoproduction of vector mesons can be used to constrain the behaviour of the gluon distribution at small-$x$ and/or the magnitude of the nuclear effects. 
However, at the high center-of-mass energies probed in $\gamma h$ interactions at LHC, new dynamics effects associated to nonlinear corrections to the QCD dynamics are expected to be present. These effects are easily included if the process is described in the color dipole formalism, with the cross section being strongly dependent on the model for the dipole - target scattering amplitude. This alternative 
approach has been used by several authors in the last years, considering different assumptions for the meson wave function and QCD dynamics, as well as for the free parameters. Our goal in this paper was, using the color dipole formalism, to estimate the theoretical uncertainty associated to description of the QCD dynamics. We have assumed a unique model for the $J/\Psi$ wave function and considered four models for the dipole - proton scattering amplitude. We demonstrated that although these models satisfactorily 
describe the HERA data, their predictions are very distinct for the diffractive photoproduction of $J/\Psi$ in $pp/pPb/PbPb$ collisions. In particular, our results point out that the recent ALICE and LHCb data are quite well described by the new version of the bCGC model. Our main conclusion is that the color dipole formalism is able to describe these data and that future measurements can be useful to constrain the magnitude of the nonlinear effects in the QCD dynamics.

\begin{table}
\begin{center}
\begin{tabular} {||c|c|c|c||}
\hline
\hline
 & {\bf GBW}& {\bf bCGC} & {\bf bCGC NEW}   \\
\hline
{\bf $pp$} ($\sqrt{s} = 7$ TeV)& 74.0 nb  & 49.0 nb  & 59.0 nb  \\
\hline
{\bf $pp$} ($\sqrt{s} = 14$ TeV)& 113.0 nb & 71.2 nb  & 93.7 nb  \\
\hline
\hline
{\bf $pPb$} ($\sqrt{s} = 5$ TeV)& 51.3  $\mu$b & 41.0 $\mu$b  & 42.8 $\mu$b \\
\hline
\hline
{\bf $PbPb$} ($\sqrt{s} = 2.76$ TeV) & 18.2 mb  & 13.6 mb  & 11.0 mb  \\
\hline
{\bf $PbPb$} ($\sqrt{s} = 5.5$ TeV)&  33.8 mb & 24.4 mb  & 20.3 mb  \\
\hline
\hline
\end{tabular}
\end{center}
\caption{The total cross section for the diffractive photoproduction of $J/\Psi$  in $pp$, $pPb$ and $PbPb$  collisions at  LHC energies.}
\label{tab}
\end{table}

%\begin{figure}[t]
%\centerline{\psfig{file=sec_coh_pel3.eps,width=60mm}}
% \caption{Produ\c c\~ao de m\'esons em colis\~oes $\gamma p$, tentativa 3.}
%\label{fig1}
%\end{figure}

\section*{Acknowledgements}

This work was partially financed by the Brazilian funding agencies CAPES, CNPq,  FAPESP and FAPERGS.

%\begin{figure}[t]
%\centerline{\psfig{file=kkp_hkns.eps,width=100mm}}
% \caption{Illustration of the coherent process $h_1 + h_2 \rightarrow h_1 h_2 \eta_c$ (See text).}
%\label{fig1}
%\end{figure}

%%%%%%%%%%%%%%%%%%%%%%%%%%%%%%%%%%%%%%%%%%%%%%%%%%%%%%%%%%%%%%%%%%%%%%%%%%%%%%%%

%%%%%%%%%%%%%%%%%%%%%%%%%%%%%%%%%%%%%%%%%%%%%%%%%%%%%%%%%%%%%%%%%%%%%%%%%%%%%%%%

\end{document}